\documentclass[nohyper,12pt,a4paper]{JHEP3} 

\usepackage{amsmath,amssymb,cite}

\def\a{\alpha}
\def\b{\beta}

\def\d{\delta}
\def\e{\epsilon}

\def\f{\phi}

\def\g{\gamma}

\def\l{\lambda}

\def\oo{\omega}
\def\p{\pi}
\def\q{\theta}
\def\r{\rho}
\def\s{\sigma}
\def\t{\tau}

\def\x{\xi}
\def\z{\zeta}

\def\F{\Phi}

\def\Ld{\Lambda}
\def\OO{\Omega}

\def\ca{{\cal A}}

\def\cf{{\cal F}}

\def\cm{{\cal M}}

\def\co{{\cal O}}

\def\cx{{\cal X}}

\newcommand{\ti}{\tilde}

\newcommand{\pa}{\partial}

\newcommand{\nn}{\nonumber}

\def\tA{{\tilde{A}}}

\def\pa{\partial}
\def\be{\begin{equation}}
\def\ba{\begin{eqnarray}}
\def\ea{\end{eqnarray}}

\def\dt{\tilde{\d}}

\def\xh{\hat{x}}
\def\yh{\hat{y}}

\def\dt{\tilde{\d}}

\allowdisplaybreaks[4]

\title{The Seiberg-Witten Map on the Fuzzy Sphere}

\author{Jesper M\o ller Grimstrup, Thordur Jonsson, L\'arus Thorlacius\\
University of Iceland, Science Institute \\
Dunhaga 3, 107 Reykjavik, Iceland\\
E-mail: \email{jesper@raunvis.hi.is}, \email{thjons@raunvis.hi.is}, \email{lth@raunvis.hi.is} }

\abstract{We construct covariant coordinate transformations on the fuzzy sphere and utilize these to construct a covariant map from a gauge theory on the fuzzy sphere to a gauge theory on the ordinary sphere. 
We show that this construction coincides with the Seiberg-Witten map on the Moyal plane in the appropriate limit. 
The analysis takes place in the algebra and is independent of any star-product representation.
}

\preprint{RH-09-2003}

\begin{document}

\section{Introduction}

One of the surprises arising in noncommutative gauge theory is the existence 
of the so-called Seiberg-Witten (SW) map between noncommutative and commutative gauge theories. Its existence on noncommutative $\mathbb{R}^4$ was first deduced from the observation that different worldsheet regularization schemes (point-splitting vs. Pauli-Villars) lead either to a commutative or a noncommutative gauge theory in the field theory limit of open string theory \cite{Seiberg:1999vs}. Subsequently, the SW-map has been extensively studied. An especially interesting approach is set within the Kontsevich star product formalism \cite{Kontsevich:1997vb}. Here the SW-map is found to constitute a part of a map between equivalent star products \cite{Jurco:2000fb,Jurco:2000fs,Jurco:2001my,Jurco:2001kp}. In particular, this study shows that the SW-map is an integral feature of any noncommutative geometry obtained through deformation quantization of a Poisson manifold. The analysis allows in principle the construction of the SW-map of both Abelian and non-Abelian gauge theories although the non-Abelian case proves delicate.\\

On noncommutative $\mathbb{R}^n$ characterized by a constant noncommutativity parameter $\q$, the SW-map can be constructed using various other techniques. Most often, the map is studied within a star-product representation of the noncommutative algebra. Switching to a BRST-setting the map can be found perturbatively using a cohomological approach \cite{Jurco:2001rq,Cerchiai:2002ss} (See also \cite{Barnich:2003wq} and references therein for recent work set within both the BRST and Batalin-Vilkovisky formalisms). This method applies to all gauge groups. Further, the SW differential equation may be derived using an ansatz of covariant coordinate transformations \cite{Bichl:2001yf,Grimstrup:2001ja,Jackiw:2001jb}. There is also an interesting connection between gauge theory on noncommutative $\mathbb{R}^n$ and fluid dynamics \cite{Susskind:2001fb}, which allows an interpretation of the SW-map in terms of the map from the Lagrange to the Euler picture of the fluid \cite{Jackiw:2002pn}.\\

On curved noncommutative spaces less is known about the SW-map. 
The fuzzy sphere, $S^2_N$, is a simple example of a noncommutative algebra \cite{Madore:1991bw,Grosse:1995ar}. It is related to the function algebra of the sphere through a cut-off of high angular momenta and is an algebra of finite matrices. The fuzzy sphere is linked to the Moyal plane, $\mathbb{R}^2_\q$, through a special limit where a region of the fuzzy sphere is infinitely enlarged \cite{Madore:1991bw,Chu:2001xi}.
A star-product representation of the fuzzy sphere can be constructed via deformation quantization of its commutative counterpart \cite{Kishimoto:2001qe}. Although the star-product will involve infinitely many differentiations, the quantization of the noncommutativity parameter truncates the series \cite{Presnajder:1999ky}. This means that the work of Jurco et al. regarding the SW-map applies also to the fuzzy sphere. In \cite{Hayasaka:2002db} a SW-map on the fuzzy sphere was constructed to second order using cohomological techniques. However, to this end a star-product on the sphere was applied which does not truncate the function algebra and thereby represents an infinite-dimensional algebra.\\

In the present paper we are concerned with an explicit construction of the SW-map on the fuzzy sphere. We apply the method of covariant coordinate transformations and use results known from the Moyal plane combined with the Moyal limit of the fuzzy sphere to identify the correct setup on the fuzzy sphere. We first construct noncommutative conformal transformations on the fuzzy sphere and subsequently exploit these to construct the SW-map. We choose to work directly in the algebra, avoiding any use of a star-product. The constructed SW-map is a gauge-covariant map from the finite dimensional algebra $S^2_N$ into the infinite dimensional algebra of functions on $S^2$. Our construction allows straightforward computation of the deformation to arbitrary orders.\\

Let us give a short outline of the strategy applied in this paper:
\begin{itemize}
\item
In a first step, all coordinate transformations on the fuzzy sphere $S^2_N$ are determined.
These coordinate transformations do not all commute with gauge transformations of a gauge field defined on the fuzzy sphere. However, combined with an appropriate transformation of the product on $S^2_N$ they form a total transformation which commutes with gauge transformations.
\item 
In a second step, new, modified coordinate transformations are proposed. These transformations are constructed so that they commute with gauge transformations. Also, modified transformations of the product are obtained which again commute with gauge transformations.
\item
From the modified transformations of the product, tensorial transformations (rank two) are constructed which again commute with gauge transformations.
\item
These tensor transformations can be interpreted as a shift in the noncommutativity parameter. Combined with a map of elements of the fuzzy sphere into the function algebra of the ordinary sphere they are used to construct the Seiberg-Witten map as a formal Taylor expansion. \\
\end{itemize}


\section{The Seiberg-Witten map}\label{11}

Briefly stated, a SW-map is a map between gauge theories defined on a noncommutative space and gauge theories on the corresponding commutative space which maps gauge equivalent classes onto gauge equivalent classes. The map is constructed to permit a {\it perturbation} of the commutative theory in the noncommutativity parameter which characterizes the noncommutative algebra.

More specifically, let a general noncommutative algebra $\ca_\q$ be characterized by a noncommutativity parameter $\q$.  For $\q=0$ the algebra $\ca_\q$ coincides with an algebra $\cm$ of functions on a manifold. We assume that a gauge theory can be formulated on $\ca_\q$ involving a gauge field $A$ and possibly matter fields $\Psi$. We denote by $\d_{\Ld}$ the (infinitesimal) noncommutative gauge transformation involving the gauge parameter $\Ld$. For $\q=0$ the corresponding gauge field, matter field, gauge parameter and gauge transformation are denoted by $a$, $\psi$, $\l$ and $\d_\l$. By a SW-map we understand a map 
\ba
\r: \ca_\q \rightarrow \cm
\ea
so that: 
\begin{enumerate}
\item
The functions $\r(A)$ and $\r(\Psi)$ takes the form of an expansion of the noncommutative fields in terms of the deformation parameter $\q$ and the corresponding commutative field
\ba
\r(A_i)=a_i + \q\cx_{i}[a]+\co(\q^2)\;,
\qquad\r(\Psi)=\psi + \q\cx[a,\psi] + \co(\q^2)\;,
\ea
where $\cx_i[a]$ and $\cx[a,\psi]$ are functions of the commutative fields. Usually, one omits the $'\r\;'$ and simply writes $A(a,\q)$, etc. It is, however, important to bear in mind that the SW-map is a map from the noncommutative to the commutative algebra\footnote{In particular, for finite dimensional algebras, such as the fuzzy sphere, one can in principle map the noncommutative algebra into the commutative one without loss of information but the reverse is clearly impossible.}. 
\item
The map respects the gauge symmetry in the sense that the {\it gauge equivalence condition} must hold:
\ba
A(a,\q) + \d_{\Ld}A(a,\q) =A(a+\d_{\l^\prime} a,\q)\;,\label{gec}
\ea
where $\l^\prime$ is a commutative gauge parameter related to $\Ld$ and $\l$. For the matter fields the equivalence condition reads
\ba
\Psi(\psi,a,\q) +\d_{\Ld}\Psi(\psi,a,\q) =\Psi(\psi+\d_{\l^\prime}\psi,a+\d_{\l^\prime} a,\q)\;.
\ea
\end{enumerate}
The SW-map allows a formal expansion of the action of the commutative theory \cite{Madore:2000en,Jurco:2000ja,Bichl:2001gu}\footnote{see also \cite{Calmet:2001na,Aschieri:2002mc} for an expansion of the noncommutative Standard Model and GUT theories.} in the noncommutativity parameter $\q$ 
\ba
S[a,\psi,\q]=S_0[a,\psi] + \sum_{n=1}^{\infty}\q^n S_n[a,\psi]\;.\label{gec2}
\ea
Such an expansion -- on noncommutative $\mathbb{R}^n_\q$ in particular -- is, in its quantized version, free of problems arising in the non-expanded theory due to UV-IR mixing \cite{Minwalla:1999px} (see also \cite{Grosse:2003nw,Grosse:2003aj} for some recent developments).  
However, the expansion in $\q$ naturally involves a coupling constant of negative dimension and thus jeopardizes renormalizability. This question was studied in \cite{Bichl:2001nf,Bichl:2001cq} for the noncommutative Maxwell theory and in \cite{Wulkenhaar:2001sq,Grimstrup:2002af} it was shown that $\q$-expanded models must involve all possible terms in the action linear in $\q$ permitted by power-counting. At higher orders in $\q$ things become more involved. The conclusion seems to be that theories expanded in $\q$ should be regarded as effective theories only \cite{Grimstrup:2002af}.

\section{The Fuzzy sphere and the Moyal plane}\label{2}

\subsection{The fuzzy sphere}

The fuzzy sphere \cite{Madore:1991bw} is one of the simplest noncommutative spaces. As an algebra of matrices, it can be regarded an approximation to the commutative sphere obtained by cutting off the function algebra on the sphere. If we denote by $x_i$, $i\in\{1,2,3\}$, Cartesian coordinates in $\mathbb{R}^3$, the sphere is defined by the constraint
\ba
x_i x_i = R^2 \label{con1}\;,
\ea
where $R$ is the radius of the sphere. We use Einstein's convention of summation over repeated indices and write indices as subscripts. By $\hat{x}_i$, $i\in\{1,2,3\}$, we denote the Hermitian generators of the fuzzy sphere which satisfy 
\ba
\hat{x}_i \hat{x}_i = R^2\;, \label{const2}
\ea
and
\ba
[\hat{x}_i,\hat{x}_j]= {\rm{i}}\a\e_{ijk}\hat{x}_k\;,
\label{thealgebradu}
\ea
where $\a$ is a real parameter. A simple rescaling leads to the $su(2)$ generators, $J_i \equiv \xh_i/\a $, which fulfill the usual commutation relation
\ba
[J_i,J_j]= {\rm{i}}\e_{ijk}J_k\;.
\ea
Irreducible representations of the $su(2)$ Lie algebra are labeled by $j$ and have a quantized Casimir invariant, $J_i J_i = j(j+1)$. This combined with the constraint (\ref{const2}) leads to the relation 
\ba
R^2 = \a^2 j(j+1) \label{relationsship}\;,
\ea
between the radius $R$ and the noncommutativity parameter $\a$ which means that $\a$ is a discrete quantity when the radius is kept fixed.
We shall use the eigenstates $|m\rangle $ of the $su(2)$ generator $J_3$, $J_3|m\rangle =m|m\rangle$, as a basis
with $m=-j, -j+1, \ldots , j$. 


Classical scalar field theory on the fuzzy sphere \cite{Grosse:1995ar} is characterized by the action functional
\ba
S[\F_N] = \frac{4\p}{N+1}\mbox{Tr}\big\{[J_i,\F_N][\F_N,J_i]+ R^2V(\F_N)\big\}\;, \qquad N=2j \;,\label{ac1}
\ea
where $V(\F_N)$ is the potential and $\F_N$ a hermitian $(N+1)\times(N+1)$ matrix.

There are several alternative approaches to the definition of a gauge theory on the fuzzy sphere \cite{Klimcik:1997mg,Grosse:1998da,Carow-Watamura:1998jn,Balachandran:1999hx,Steinacker:2003sd}. Since our analysis of the Seiberg-Witten map is in fact independent of the action -- the only inputs necessary are (infinitesimal) gauge transformations and the 'spherical constraint' (see below) -- we believe that it is independent of the actual setup of gauge theory on $S^2_N$. Nevertheless, for completeness let us write down a possible action describing a gauge theory. We restrict ourselves to the Abelian $U(1)$ case\footnote{For a treatment of the non-Abelian case see for example \cite{Steinacker:2003sd}.} and write
the action functional as \cite{Carow-Watamura:1998jn}
\ba
S[A]= \frac{4\p R^2}{N +1}\mbox{Tr}\{\cf_{ij} \cf_{ij} \} \label{act}\;.\label{ac2}
\ea
Here, $\cf_{ij}$ is the noncommutative field strength given by
\ba
\cf_{ij} &=& {\rm{i}}L_i A_{j}-{\rm{i}}L_{j} A_{i} +\e_{ijk}A_{k} +{\rm{i}}R[A_i,A_{j}]\;,
\ea
where $L_i$ is the Lie derivative 
\ba
L_i\ast = \frac{1}{\a}[\xh_i,\ast] = [J_i,\ast]\;.\label{derivation}
\ea
The action (\ref{act}) is invariant under $U(1)$ gauge transformations\footnote{We adopt a notation slightly different from that of \cite{Carow-Watamura:1998jn}: We introduce a factor $R$ in order to obtain the correct gauge field in the Moyal limit. Further, we choose to study scalar fields in the adjoint representation.}
\ba
A_i\rightarrow A_i^\prime = \frac{1}{R} u L_k u^\ast + u A_k u^\ast \label{gt}\;,
\ea
where $u\in S^2_N, uu^\ast=1$. The transformation (\ref{gt}) has the infinitesimal form 
\ba
A_i \rightarrow A_i^\prime = A_i +\d_\Ld A_i\;, \qquad \d_\Ld A_i = \frac{1}{\a R}[\xh_i,\Ld] + {\rm{i}}[A_i,\Ld] \label{hik}\;,
\ea
where $\Ld$ is an infinitesimal element of the algebra.
Gauge theory on the classical sphere involves three components $a_i$ as well as the constraint $x_i a_i =0$. On $S^2_N$ the constraint can be generalized to
\ba
\xh_i A_i +  A_i \xh_i + \a R A_i A_i
=0\;.\label{hovsa}
\ea
A complex scalar field theory on the fuzzy sphere can be coupled to a gauge theory using the minimal coupling
\ba
[\xh_i,\ast]\rightarrow [X_i,\ast]\;,
\ea
where 
\ba
X_i = \xh_i + \a R A_i \;, \label{covariantcoordinate}
\ea
which leads to the action
\ba
S[\F_N,A] = \frac{4\p}{\a^2(N+1)}\mbox{Tr}\big\{[X_i,\F_N][\F_N,X_i]+ \a^2R^2V(\F_N)\big\}\;. \label{ac3}
\ea
$X_i$ is sometimes referred to as a covariant coordinate \cite{Madore:2000en}, since it has the property
\ba
\d_\Ld (X_i \F_N) = {\rm{i}}[\Ld,X_i\F_N]\;,
\ea
provided the field $\F_N$ transforms in the adjoint representation:
\ba
\d_\Ld\F_N={\rm{i}}[\Ld,\F_N] \;. 
\ea
Further, the covariant coordinates satisfy
\ba
[X_i,X_j]= -{\rm{i}}\a^2 R \cf_{ij} + {\rm{i}}\a\e_{ijk}X_k\;.
\ea
The fuzzy constraint (\ref{hovsa}) has in terms of covariant coordinates the suggestive form
\ba
X_i X_i = R^2 \;,\label{correctconstraint}
\ea
which is the covariant version of the spherical constraint (\ref{const2}).

\subsection{The Moyal plane}\label{3}

The Moyal plane is generated by two 'coordinates'; hermitian operators $\hat{y}_i$, $i\in \{1,2\}$ fulfilling the algebra
\ba
[\yh_i,\yh_j]= {\rm{i}}\q\e_{ij}\;,
\ea
where $\q$ is constant, real parameter of power-counting dimension $-2$. We define
annihilation and creation operators by
\ba
&&a = \frac{1}{\sqrt{2\q}}(\yh_1 + {\rm{i}}\yh_2) \;,\qquad a^\dagger=\frac{1}{\sqrt{2\q}} (\yh_1 - {\rm{i}}\yh_2)\;,
\ea
and use the harmonic oscillator basis
\ba
|n\rangle =\frac{(a^\dagger)^n}{\sqrt{n!}}|0\rangle\;,\quad a|n\rangle = \sqrt{n} |n-1\rangle \;,\quad a^\dagger|i\rangle = \sqrt{n+1} |n+1\rangle\;,
\ea
to expand noncommutative fields such as the scalar field $\F^{\mbox{\tiny m}}$, $\F^{\mbox{\tiny m}}_{ij} \equiv \langle i|\F^{\mbox{\tiny m}}|j\rangle$.
The superscript${\;}^{\mbox{\tiny m}}$ indicates objects in the Moyal plane.

\section{Limits of the Fuzzy sphere}\label{4}

There are two obvious limits of the fuzzy sphere: The classical limit given by $\a\rightarrow 0$, where the algebra becomes commutative, and the Moyal limit given by $\a R \rightarrow \q$ together with $j\rightarrow\infty$, where the 'north pole' is blown up in such a way that $\xh_3$ enters the center of the algebra, which then reduces to the algebra of the Moyal plane.

\subsection{The classical limit $S^2$}

In the classical limit we let the angular momentum cut-off $j$ approach infinity while $R$ is kept fixed. The relation (\ref{relationsship}) implies that $\a$ vanishes and the algebra becomes commutative $[x_i,x_j]=0$. Note that the Lie derivative (\ref{derivation}) converges towards the classical generator of angular momentum, $J^{\mbox{\tiny cl}}_j= -{\rm{i}}\e_{jkl}x_k\pa_l$ which is seen by writing an element of $S^2_N$ as an expansion in normal ordered generators $\xh_i$ and calculating a commutator with $\xh_j$. Consequently, the actions (\ref{ac1}), (\ref{ac2}) and (\ref{ac3}) converge towards their classical counterparts\footnote{The convergence of matrix algebras to the commutative sphere was carefully studied in \cite{Rieffel:2001qw}. Further, the convergence of ``fuzzy actions'' in the commutative limit was studied in \cite{Balachandran:2000wp}.}
\ba
S[\F^{\mbox{\tiny cl}}] =\int\Big(\big(J^{\mbox{\tiny cl}}_i\F^{\mbox{\tiny cl}}\big)\big(J^{\mbox{\tiny cl}}_i\F^{\mbox{\tiny cl}}\big)+ R^2 V(\F^{\mbox{\tiny cl}})\Big)\;, \quad \mbox{etc.}\;\ldots\;,
\ea
since the integral on the fuzzy sphere
\ba
 \frac{4\p}{N+1}\mbox{Tr}\big\{\x_N  \big\} \;,\quad\x_N\in S^2_N\;,
\ea
agrees with the integral on $S^2$ for $N\rightarrow\infty$:
\ba
\frac{4\p}{N+1}\mbox{Tr}\big\{\x_N  \big\}\rightarrow \int_{S^2} \x\;,\qquad \mbox{for}\quad \x_N \rightarrow \x\;.
\ea

\subsection{The Moyal limit $\mathbb{R}^2_\q$}

Roughly speaking, if an infinitesimal region around the `north pole' is blown up in a way so that $\a\xh_3$ approaches a constant denoted $\q$, $\xh_3$ should enter the center of the algebra and we would expect to end up with a plane equipped with a constant commutator relation $[\xh_1,\xh_2]={\rm{i}}\q$, {\it i.e.} the Moyal plane. In this limit, everything away from the north pole approaches infinity in the Moyal plane.

To be precise we take the large $j$ and large $R$ limit so that 
\ba
j\rightarrow\infty\;,\qquad R^2 = \frac{j\q}{4}\rightarrow\infty \qquad\mbox{keeping $\q$ fixed}\;,
\ea
and consider the weak convergence of matrices $\F_N$ in the fuzzy sphere towards operators $\F^{\mbox{\tiny m}}$ in the Moyal plane by zooming in on the `top-left' corner of the matrices:
\ba
\lim_{j \rightarrow \infty} \langle j-k |\F_N| j-k^\prime\rangle = \langle k| \F^{\mbox{\tiny m}} | k^\prime\rangle\;.\label{hhh}
\ea
In the sequel we will denote this limit by `$\rightsquigarrow$' and will in general leave out matrix indices. Thus, (\ref{hhh}) is written
\ba
\F_N \rightsquigarrow \F^{\mbox{\tiny m}}\;.
\ea
Let us investigate this limit in some detail. Clearly we find that $\xh_3$ diverges in the Moyal limit
\ba
\xh_3 \sim \sqrt{j}\rightsquigarrow \infty\;,
\ea
whereas 
\ba
\a\xh_3 \rightsquigarrow \q\;,
\ea
as expected. Also, we find that a commutator with $\xh_3$ vanishes
\ba
[\xh_3,\cdot]\rightsquigarrow 0 \;,
\ea
which shows that $\xh_3$ enters the center of the algebra in the Moyal limit. 
For completeness, we also note that 
\ba
 \a [\F_N ,J_+] \rightsquigarrow \sqrt{2\q} [\F^{\mbox{\tiny m}},a ] \;,\qquad \a [\F_N ,J_-] \rightsquigarrow \sqrt{2\q} [\F^{\mbox{\tiny m}},a^\dagger ]\;,\label{58}
\ea
where $J_\pm = J_1 \pm {\rm{i}}J_2$.

Let us look at the Moyal limit of a gauge theory on the fuzzy sphere. In particular, we are interested in the constraint (\ref{correctconstraint}). To obtain the correct gauge theory in the Moyal limit we divide out a factor $R$ and write the constraint as
\ba
\a\big(\{\xh_i,A_i \}+\a R A_i A_i  \big)=0\;.
\ea
All terms in the constraint vanish due to the overall factor $\a$ (which scales as $j^{-1/2}$) except terms involving $\xh_3$ since this scales as $j^{1/2}$. In the Moyal limit, therefore, the gauge constraint forces $A_3$ to scale to zero
\ba
A_3\rightsquigarrow 0\;,
\ea 
whereas we find no constraint on the other components
\ba
A_{1,2}\rightsquigarrow A^{\mbox{\tiny m}}_{1,2}\;.
\ea

Let us close this section with some general considerations. Given a sequence of fields on fuzzy spheres, $\{\F_N\}$, $N\in \{1,\ldots \}$, which converges weakly towards a field $\F^m$ on the Moyal plane, and given that $\F_N$ fulfills the equation of motion on the respective fuzzy sphere, then it is clear from (\ref{58}) that the field $\F^m$ will satisfy the equations of motion in the Moyal plane since the equations of motions themselves converge weakly. A specific example is provided by the soliton solution on the fuzzy sphere given in \cite{Austing:2002ih} which was shown to converge weakly towards a soliton solution of the Moyal plane.

\section{The Seiberg-Witten map on the Moyal plane}\label{5}

Since we have a good understanding of the SW-map in the Moyal plane, we might hope to gain information about a SW-map on the fuzzy sphere through the weak convergence described in the previous section. It seems plausible that a SW-map on the fuzzy sphere should converge weakly towards its counterpart in the Moyal plane. Let us therefore first give a short review of a suitable derivation of the SW-map on the Moyal plane (See \cite{jesper} for details.). In the following sections we apply the approach to the fuzzy sphere.

In the plane, a classical, infinitesimal dilation has the form
\ba
\d_{\e}\F = \e x^\a \pa_\a \F\;, \label{cl}
\ea
where $\e\ll 1$ and $\F$ denotes a classical field. A noncommutative version of this could have the form 
\ba
\d_{\e}\F^{\mbox{\tiny m}} &=& \frac{{\rm{i}}\e}{2\q}\e_{\a\b}\{\yh_\a,[\yh_\b,\F^{\mbox{\tiny m}}]\} \label{di}\;,
\ea
since it converges towards (\ref{cl}) in the commutative limit and fulfills an appropriate conformal algebra. 
An infinitesimal gauge transformation is defined by
\ba
\d^g_\l \F^{\mbox{\tiny m}} = {\rm{i}}[\l,\F^{\mbox{\tiny m}}]\;,
\ea
where $\l$ is an infinitesimal element of the algebra, $\l\in\mathbb{R}^2_\q$. The commutator of $\d_\e$ and a gauge transformation is easily calculated 
\ba
[\d^g_\l, \d_{\e}] \F^{\mbox{\tiny m}} &=&
\d^g_{\l^\prime}\F^{\mbox{\tiny m}} -  \frac{\e}{\q}\e_{\a\b}\{[\yh_\b,\l],[\yh_\a,\F^{\mbox{\tiny m}}] \} \;, \label{portishead}
\ea
with
\ba
\l^\prime = \frac{{\rm{i}}\e}{2\q}\e_{\a\b}\{\yh_\a,[\yh_\b,\l]\} - \d_\e\l\;.
\ea
Since gauge transformations and the dilation (\ref{di}) do not form a closed algebra one postulates new noncommutative coordinate transformations\footnote{One can give a physical argument as to why covariant active transformations should exist \cite{Bichl:2001yf}: Given an observable $\co$ of a noncommutative field theory which has a conformal field theory as its commutative limit. We write $\d^{g}_\l\co =0$ and consider the transformed observable, $\co^\prime = \d_{\e}\co$. $\co^\prime$ does not equal $\co$ because the noncommutativity breaks conformal invariance. It should, however, still be an observable, $\d^g_\l\co^\prime =0$. This implies that the commutator $\big[\d^g_\l,\d_{\e}\big]$ must be another symmetry transformation of the theory.}
\ba
\tilde{\d}_{\e}\F^{\mbox{\tiny m}}= \d_\e\F^{\mbox{\tiny m}} + \cx^{\mbox{\tiny m}}\;,\label{oneone}
\ea
which satisfy the algebra
\ba
[\d^g_\l, \tilde{\d}_{\e}] \F^{\mbox{\tiny m}} &=&
\d^g_{\l^\prime}\F^{\mbox{\tiny m}} \;,\label{twotwo}
\ea
where $\l^\prime$ is a (possibly field dependent) gauge parameter. Basically, $\cx^{\mbox{\tiny m}}$ is the missing piece needed to write the coordinate transformation $\d_{\e}$ in a covariant way. Inserting (\ref{oneone}) into (\ref{twotwo}) one eventually finds\footnote{Usually, the gauge field on the Moyal plane is defined as $A^{\prime\mbox{\tiny m}}_i =\e_{ij}A^{\mbox{\tiny m}}_j$. With this convention $\cx^{\mbox{\tiny m}}$ in (\ref{kathrin}) agrees with results found elsewhere in the literature.}
\ba
\cx^{\mbox{\tiny m}} =   \e \q\e_{jk}\Big(
\{A^{\mbox{\tiny m}}_j,\frac{{\rm{i}}}{\q}[\yh_k,\F^{\mbox{\tiny m}}]\}
+\frac{{\rm{i}} }{2}\{ A^{\mbox{\tiny m}}_j,[A^{\mbox{\tiny m}}_k,\F^{\mbox{\tiny m}}]\}\Big)
+\OO^{\mbox{\tiny m}} \label{kathrin}\;,
\ea
where $\OO^{\mbox{\tiny m}}$ is an arbitrary covariant term proportional to $\q$. If we write $\q_{ij}=\q\e_{ij}$ (\ref{kathrin})
can be written as (we set $\OO^{\mbox{\tiny m}} $ to zero)
\ba
\cx^{\mbox{\tiny m}} := \d_\q \F^{\mbox{\tiny m}} = \d\q_{ij}\cx_{ij}^{\mbox{\tiny m}} \;, \label{fiur}
\ea
where $\cx_{ij}$ can be read off from (\ref{kathrin}). The transformation $\d_\q$ is a modified shift of $\q$ (and thus also acts on the product) and one can show that the algebra of $\d_\q$ and gauge transformations closes
\ba
[\d_\q,\d^g_\l]\F^{\mbox{\tiny m}} = \d^g_{\l^\prime}\F^{\mbox{\tiny m}}\;,\label{mormor}
\ea
which is a formulation of the gauge equivalence condition (\ref{gec2}). Upon expanding (\ref{mormor}) around $\q=0$ via a formal Taylor expansion one eventually constructs the SW-map. 

Further, equation (\ref{fiur}) may be interpreted as a differential equation, the SW differential equation, which in a star-product setting reads
\ba
\frac{d \F^{\mbox{\tiny m}}}{d \q_{ij}} =  \cx_{ij}^{\mbox{\tiny m}} \;,\label{jaujau}
\ea
where all products in (\ref{jaujau}) have been replaced with appropriate star-products. Analogously one finds differential equations for the gauge field as well as for any matter field(s) transforming under gauge transformations. This approach also works for non-Abelian gauge groups \cite{Bichl:2001yf}.\\

In the commutative case conformal transformations of the gauge field lead to an energy-momentum tensor which is not gauge invariant and thus unphysical. This deficit is corrected by adding a field-dependent gauge transformation to the conformal transformation \cite{Jackiw:ar}. The algebra of conformal transformations then closes only up to gauge transformations. The important observation is that the commutative conformal transformations of the gauge field may be written as a gauge covariant part and a part which forms a gauge transformation. Subtracting this gauge transformation leads to gauge covariant transformations. In the noncommutative case this does not work. Due to the noncommutative product the (initial) conformal transformation of the gauge-field cannot be rewritten as a covariant part and a gauge transformation. The missing part is exactly the piece of information which can be interpreted as the $\q$-dependence of the gauge field and which is given by the Seiberg-Witten differential equation (see also \cite{Jackiw:2001jb}).

\section{Infinitesimal coordinate transformations on $S^2$ and $S^2_N$}\label{6}

On the classical sphere a dilation similar to (\ref{cl}) would violate the constraint (\ref{con1}). In general, an infinitesimal coordinate transformation
\ba
x_i \rightarrow x_i^\prime = x_i - f_i\;, \quad f_i \ll 1\;,
\ea
on $S^2$ must fulfill the requirement
\ba
f_i x_i = 0\;,
\ea
which requires $f_i$ to be of the form
\ba
f_i = \oo_{ij}x_j + \frac{\e_j}{R} (x_i x_j - \d_{ij}x^2)\;,\label{fd}
\ea
where $\oo_{ij}= -\oo_{ji}$ and $\e_i$ are infinitesimal parameters. It turns out that rotations on the fuzzy sphere commute with gauge transformations -- as is the case in the Moyal plane -- and thus do not interest us (see next section for details). In the following we are therefore solely interested in the infinitesimal transformations characterized by $\e_i$ which we shall here call {\it spherical dilations}. The commutator of two spherical dilations gives ($\d_{\e} x_i \equiv f_i$)
\ba
[\d_{\e_1},\d_{\e_2}]x_i = R^2((\e_1)_k (\e_2)_i - (\e_1)_i (\e_2)_k)x_k\;,
\ea
and thus closes on a rotation.
A scalar field $\F$ on the sphere transforms under spherical dilations according to 
\ba
\d_{\e}\F &=& 
 {\rm{i}} \frac{\e_j}{R}\e_{jkl} x_k  J^{\mbox{\tiny cl}}_l \F \label{lalala}\;.
\ea
Further, a gauge field $a_i$ on the sphere transforms like 
\ba
\d_{\e} a_i &=& f_j\pa_j a_i + \pa_i f_j a_j
\nn\\&=& {\rm{i}} \frac{\e_j}{R} \e_{jkl} x_k  J^{\mbox{\tiny cl}}_l a_i + \frac{\e_j}{R} \big( x_j a_i+ \d_{ij}x_k a_k -2 x_i a_j  \big)\;.
\ea
Notice that the transformations defined by (\ref{fd}) reduce to dilations on the plane when one zooms in on the north pole and chooses $\e_1,\e_2=0$ and $\e_3=\e\ll1$.

We now formulate a noncommutative version of these transformations on the fuzzy sphere. It turns out that the transformation of a scalar field $\F_N$ according to
\ba
\d_{\e}\F_N = \frac{{\rm{i}}\e_i}{2\a R}\e_{ijk}\{\xh_j,[\xh_k,\F_N]\}\;,\label{fuzzz}
\ea 
gives the correct coordinate transformations in the commutative limit:
\ba
\d_{\e}\F_N \stackrel{j\rightarrow\infty}{\longrightarrow} \d_{\e}\F\;.
\ea
For the gauge field on the fuzzy sphere the transformations read
\ba
\d_{\e} A_i = \frac{{\rm{i}}\e_j}{2R\a}\e_{jkl}\{\xh_k,[\xh_l,A_i]\} + \frac{\e_j}{2R}\Big(\{ \xh_k,A_k\}\d_{ij} + \{\xh_j,A_i\} - 2 \{ \xh_i,A_j  \}\Big)\;.
\ea
We denote these transformations {\it fuzzy dilations}.  Furthermore, we shall refer to this kind of noncommutative transformations as 'active' transformations as opposed to 'passive' transformations which we will encounter in the next section. 
It is easy to show that fuzzy dilations of fields on the fuzzy sphere converge weakly to dilations of fields on the Moyal plane
\ba
\d_{\e} \F_N \rightsquigarrow \d_{\e} \F^{\mbox{\tiny m}}  \;, \qquad  \d_{\e} A_i \rightsquigarrow \d_{\e} A_i^{\mbox{\tiny m}}     \;.
\ea

\section{Covariant coordinate transformations on $S^2_N$}\label{7}

The idea is to identify (infinitesimal) noncommutative coordinate transformations of the fields which do not commute with gauge transformations. Since a coordinate is in general not a covariant object on a noncommutative space it is hardly surprising that coordinate transformations fail to form a closed algebra with gauge transformations.

\subsection{The complex scalar field}

Let us first consider infinitesimal rotations of the scalar field
\ba
\d_b \F_N = \frac{{\rm{i}}}{\a}b_i [\xh_i,\F_N]\;,
\ea
where the $b_i$'s are infinitesimal. The commutator of a gauge transformation $\d^g_\Ld$ and a rotation is
\ba
[\d_b,\d^g_\Ld]\F_N &=& \d^g_{\Ld_b}\F_N\;,\qquad \Ld_b = \d_b\Ld - \frac{{\rm{i}}}{\a}b_i[\xh_i,\Ld]\;.
\ea
Since the algebra closes we conclude that rotations are not suitable coordinate transformations for our approach. 

Next, we turn our attention to the fuzzy dilations defined by (\ref{fuzzz}) and calculate the
commutator with a gauge transformation  
\ba
[\d_{\e},\d^g_\Ld]\F_N &=&\d^g_{\bar{\Ld}}\F_N  - \frac{\e_i}{R\a}\e_{ijk}\big\{[\xh_j,\Ld],[\xh_k,\F_N]\big\}\;, 
\label{sunshine}
\ea
with
\ba
\bar{\Ld} =  \frac{{\rm{i}}\e_i}{2\a R}\e_{ijk}\{\xh_j,[\xh_k,\Ld]\} - \d_{\e}\Ld\;.
\ea
The result (\ref{sunshine}) is the sort of relation we are looking for: The algebra fails to close. Notice that the last term on the right hand side vanishes in the commutative limit as it should.
If we introduce a transformation of the noncommutative product
\ba
\d_{\e}^{\mbox{\tiny prod}} (\x_N\z_N) &=&\frac{\e_i}{R\a}\e_{ijk}[\xh_j,\x_N][\xh_k,\z_N] +(\d_{\e}^{\mbox{\tiny prod}}\x_N)\z_N+\x_N(\d_{\e}^{\mbox{\tiny prod}}\z_N) \;,\qquad \x_N,\z_N\in S^2_N\;,
\nn\\
\d_{\e}^{\mbox{\tiny prod}} \xh_i &=& 0\;,
\label{tpr}
\ea
which compensates for the noncommutativity of the algebra, and if we for now assume that the scalar field does not transform under this transformation
\ba
\d_{\e}^{\mbox{\tiny prod}}\F_N =0\;,
\ea
then the combined transformation  
\ba
\d^{\mbox{\tiny tot}}_{\e} = \d_{\e}+\d_{\e}^{\mbox{\tiny prod}}\;,\label{tott}
\ea
will form a closed algebra with gauge transformations
\ba
[\d^{\mbox{\tiny tot}}_{\e},\d^g_\Ld]\F_N &=& \d^g_{\bar{\Ld}}\F_N\;.\label{KL}
\ea
The gauge parameter $\bar{\Ld}$ now reads
\ba
\bar{\Ld} =  \frac{{\rm{i}}\e_i}{2\a R}\e_{ijk}\{\xh_j,[\xh_k,\Ld]\} - \d^{\mbox{\tiny tot}}_a\Ld\;.
\ea

In ordinary field theory there is no difference between active and passive coordinate transformations; active transformations acts on the fields whereas passive transformations act on the coordinate system in an inverse manner. In a noncommutative setting the two type of coordinate transformations are no longer equivalent. Due to the dimensional parameter $\a$ in the algebra (\ref{thealgebradu}) the noncommutative product plays the role of a background field\footnote{in a star-product representation this becomes apparent since the star-product involves the parameter $\a$.}. Therefore, by active transformations we understand transformations acting only on the fields. Passive transformations acts on both the fields and the product.
In this sense (\ref{tott}) is a passive coordinate transformation and
equation (\ref{KL}) shows that we can construct covariant passive transformations. We would, however, like to construct covariant active coordinate transformations.
Let us therefore define modified fuzzy dilations by introducing a -- yet to be determined -- additional term $\cx$
\ba
\dt_{\e}\F_N &=& \frac{{\rm{i}}\e_i}{2R\a}\e_{ijk}\{\xh_j,[\xh_k,\F_N]\} + \cx\;, 
\\
\dt_{\e} (\x_N\z_N) &=&  0\;, \qquad \x_N,\z_N\in S^2_N\;,
\ea
(the second expression states that the product does not transform under active transformations) as well as a modified transformation of the product
\ba
\dt_{\e}^{\mbox{\tiny prod}}\F_N &=& - \cx  \;, 
\\
\dt_{\e}^{\mbox{\tiny prod}}(\x_N\z_N) &=& \frac{\e_i}{R\a}\e_{ijk}[\xh_j,\x_N][\xh_k,\z_N]+(\d_{\e}^{\mbox{\tiny prod}}\x_N)\z_N+\x_N(\d_{\e}^{\mbox{\tiny prod}}\z_N)\;, 
\ea
for all $\x_N,\z_N\in S^2_N$, so that the combined transformations still add up to the passive transformation (\ref{tott})
\ba
\d^{\mbox{\tiny tot}}_{\e} =\dt_{\e}+\dt_{\e}^{\mbox{\tiny prod}} \;.
\ea
The modification term $\cx$ is to be constructed so that it provides us with a closed combined algebra of coordinate transformations and gauge transformations
\ba
[\dt_{\e},\d^g_\Ld]\F_N &=& \d^g_{\Ld^\prime}\F_N \;,\qquad[\dt^{\mbox{\tiny prod}}_a,\d^g_\Ld]\F_N = \d^g_{\Ld^{\prime\prime}}\F_N \label{closingalgebra}\;,
\ea
with
\ba
\Ld^\prime + \Ld^{\prime\prime} = \bar{\Ld}\;.
\ea
In (\ref{closingalgebra}) the second expression follows from the first using (\ref{KL}). The quantity $\cx$ should vanish in the commutative limit
\ba
\cx \stackrel{j\rightarrow\infty}{\longrightarrow}0\;.
\ea
On this basis we wish to determine $\cx$. The most general solution to (\ref{closingalgebra}) has the form
\ba
\tilde{\d}_\e =\OO + {\rm{i}}[\ti{\Ld},\F_N] \;,\label{INIT}
\ea
where $\OO$ is a covariant object and $\ti{\Ld}$ some gauge parameter, possibly field dependent. 
We find (see the appendices for details)
\ba
\ti{\Ld} =\frac{\e_i}{2}\e_{ijk}\{\xh_j,A_k\} \label{rel1} \;,
\ea
 and 
\ba
\OO  =\OO^\prime +\frac{{\rm{i}}\e_i}{2R\a}\e_{ijk}\{X_j,[X_k,\F_N]\}\;,
\ea
where $X_i$ is the covariant coordinate and $\OO^\prime $ is an arbitrary covariant term that must vanish in the commutative limit. In the following we set $\OO_a^\prime$ to zero. Finally we find
\ba
\cx &=& \a R  \e_i\e_{ijk}\cx_{jk}   \;,\label{lalalala}
\ea  
with 
\ba
\cx_{jk} &=&  \frac{{\rm{i}}}{R\a}\{A_j,[\xh_k,\F_N]\}
+ \frac{{\rm{i}}}{2}\{A_j,[A_k,\F_N]\}  \;,\label{lalalaja}
\ea  
which should be viewed as antisymmetrized in the indices $j$ and $k$.

Let us take the Moyal limit of the expression (\ref{lalalala}). We find that only the component with $\e_3\equiv\e$ yields a non-vanishing result and that (\ref{lalalala}) approaches the corresponding quantity from the Moyal plane (\ref{kathrin})
\ba
\cx &\rightsquigarrow &  \cx^{\mbox{\tiny m}}\;.
\ea 
This shows that we have identified an object on the fuzzy sphere which corresponds to the term (\ref{kathrin}) related to the SW differential equation on the Moyal plane.

Writing $\a R\e_i\e_{ijk}\equiv-\ti{\d}_\e^{\mbox{\tiny prod}} \q_{ij}$ relation (\ref{lalalala}) takes the form
\ba
\ti{\d}_\e^{\mbox{\tiny prod}}\F_N=  \big(\ti{\d}_\e^{\mbox{\tiny prod}} \q_{ij}\big)\cx_{ij}\;,
\ea
which can be interpreted as a the Seiberg-Witten differential equation on the fuzzy sphere. Here, however, it takes the form of a difference equation since the noncommutativity parameter $\a$ is not a smoothly deformable quantity.
It would be interesting to calculate the algebra of the transformations $\tilde{\d}_\e$ (or $\ti{\d}^{\mbox{\tiny prod}}$). However, since we do not need this for our purpose we leave it out.

Let us finally find the gauge parameters $\Ld^\prime$ and $\Ld^{\prime\prime}$ in (\ref{closingalgebra}). Using 
\ba
[\d^g_{\Ld_1},\d^g_{\Ld_2}]= \d^g_{\Ld_3}\;,\quad \Ld_3 = [\Ld_1,\Ld_2] + \d^g_{\Ld_1}\Ld_2 -\d^g_{\Ld_2}\Ld_1 \;,
\ea
together with (\ref{rel1}) we find
\ba
\Ld^\prime (\Ld)&=& 
 \frac{{\rm{i}}\e_i}{2}\e_{ijk}\{A_k,[\Ld,\xh_j]\}+\frac{{\rm{i}}\e_i}{2\a R}\e_{ijk} \{\xh_j,[\xh_k,\Ld]\}  - \ti{\d}_a\Ld\;,
\ea
and thus
\ba
\Ld^{\prime\prime}(\Ld)= -\frac{{\rm{i}}\e_i}{2}\e_{ijk}\{A_k,[\Ld,\xh_j]\}  - \ti{\d}^{\mbox{\tiny prod}}_a\Ld \label{accord}\;.
\ea

\subsection{The gauge field}

The analysis can be repeated for the gauge field. We first 
observe that the commutator of fuzzy dilations and gauge transformations fails to close
\ba
[\d_\e,\d^g_\Ld] A_i\not= \d^g_{\Ld^\prime}A_i \;,
\ea
which again leads us to postulate modified coordinate transformations
\ba
\ti{\d}_{\e} A_i = \d_{\e} A_i + \cx_i\;,
\ea
so that
\ba
[\ti{\d}_{\e},\d^g_{\Ld} ]A_i = \d_{\Ld^\prime}A_i\;. \label{kite}
\ea
The modified transformation $\ti{\d}_\e$ is related to the modified transformation of the product $\tilde{\d}_\e^{\mbox{\tiny prod}}$ through 
\ba
\d_\e +\d_\e^{\mbox{\tiny prod}}=\ti{\d}_\e +\ti{\d}_\e^{\mbox{\tiny prod}}\;,
\ea
and it's action on the gauge field is given by
\ba
\ti{\d}_\e^{\mbox{\tiny prod}}A_i = -\cx_i\;.
\ea
The transformation satisfy the algebra
\ba
[\ti{\d}^{\mbox{\tiny prod}}_{\e},\d^g_{\Ld} ]A_i = \d_{\Ld^{\prime\prime}}A_i\;,
\ea
whenever (\ref{kite}) is fulfilled. We proceed by noting that the full solution to (\ref{kite}) consist of a gauge transformation and a covariant part
\ba
\ti{\d}_{\e} A_i = \frac{{\rm{i}}}{\a R}[X_i,\ti{\Ld}]   +  \OO_{i} \label{saddd}\;.
\ea
We refer to the appendices for details and give the results. The covariant term $\OO_i$ reads (we omit a general covariant term proportional to $\a$)
\ba
\OO_{i} =  \frac{{\rm{i}}\e_j}{2R\a}\e_{jkl}\{X_k,[X_l,X_i]\} + \frac{ \e_j}{2R}\Big(-\{X_j, X_i \}  +2\d_{ji}R^2\Big)\;,
\ea
and the gauge parameter is found to be
\ba
\ti{\Ld} =  \frac{\e_j}{2}\e_{jkl}\{\xh_k,  A_l\}\label{rel2}\;,
\ea
which coincides with the gauge parameter (\ref{rel1}) found in the treatment of the scalar field. Finally we obtain
\ba
\cx_i &=&  \a R \e_j\e_{jkl}\cx_{ikl}\label{rel3}\;,
\ea
with 
\ba
\cx_{ikl} &=& 
{\rm{i}}\{ A_k,\frac{1}{\a R}[\xh_l,A_i]\}
+ {\rm{i}} \frac{1}{2}\{ A_k,\frac{1}{\a R}[ A_l,\xh_i]\}
+ {\rm{i}} \frac{ 1}{2}\{ A_k,[ A_l,A_i]\}
\nn\\&&
-\e_{lim}\frac{1}{2\a R^2}\{A_m,\xh_k\}   +\e_{lim}\frac{1}{2 R}   \{A_k,A_m\} \;,\label{ttt}
\ea
which should again be viewed as antisymmetrized in the indices $k$ and $l$. In the Moyal limit (\ref{rel3}) coincides with the term $\cx_i^{\mbox{\tiny m}}$ related to the SW-differential equation of the gauge field:
\ba
\cx_i &\rightsquigarrow &\cx_i^{\mbox{\tiny m}}= -\d\q \e_{kl}\Big(
{\rm{i}}\{ A_k,\frac{1}{\q}[\yh_l,A^{\mbox{\tiny m}}_i]\}
+ {\rm{i}} \frac{1}{2}\{ A^{\mbox{\tiny m}}_k,\frac{1}{\q}[ A^{\mbox{\tiny m}}_l,Y_i]\} \Big)\;,
\ea
where $Y_i= \yh_i + \q A^{\mbox{\tiny m}}_i$ and $\d\q=\q \e_3$.

\section{The Seiberg-Witten map on $S^2_N$}\label{8}

Recall that the SW-map as we defined it in section \ref{11} yields a perturbative expansion of a given noncommutative theory in powers of the parameter of noncommutativity. On the fuzzy sphere the relevant parameter is $\a$ which is, however, not  smoothly deformable when $R$ is held fixed. Nevertheless, it turns out to be possible to construct a SW-map between the fuzzy sphere and the ordinary sphere.

\subsection{The Seiberg-Witten map}

We first 
note that all relevant formulas (\ref{rel1}), (\ref{lalalaja}), (\ref{rel2}) and (\ref{rel3}) display $\e_i$ in the factor $\a \e_j \e_{jkl}$ only. Since this factor lies in the center of $S^2_N$ it means that the tensorial transformation $\d_{ij}$ defined by\footnote{We use the notation $A_{[ij]}=\frac{1}{2}A_{ij}-\frac{1}{2}A_{ji}$.}
\ba
\d_{ij}\F_N &=& \cx_{[ij]} \;,\qquad 
\nn\\
\d_{ij} A_k &=& \cx_{[ij]k}      \;,\qquad
\nn\\
  \d_{ij} (\x_N\z_N) &=& \frac{1}{2R^2\a^2}\{ [\xh_i,\x_N],[\xh_j,\z_N]\}+(\d_{ij}\x_N)\z_N+\x_N(\d_{ij}\z_N)\;,\qquad \x_N,\z_N\in S^2_N\label{maa1}\;,
\nn\\\d_{ij} \xh_k &=&0\;,
\ea
fulfills
\ba
[\d_{ij},\d^g_\Ld] = \d^g_{\Ld_{ij}} \label{maa2}\;,
\ea
with 
\ba
\Ld_{ij}(\Ld)= \frac{1}{4\a R}\Big(\{A_i,[\Ld,\xh_j]\}- \{A_j,[\Ld,\xh_i]\}\Big) - \d_{ij}\Ld\;,\label{gpa}
\ea
where, according to (\ref{accord}), we have left open the possibility that the gauge parameter $\Ld$ is transformed under $\d_{ij}$. The commutator (\ref{maa2}) is the crucial relation we need to construct a SW-map on the fuzzy sphere, since it is a two-component transformation which commutes with gauge transformations. The transformation $\d_{ij}$ resembles a shift in the commutator $[\xh_i,\xh_j]$. Using (\ref{maa2}) we can calculate the commutator of $n$ $\d_{ij}$'s with a gauge transformation. For $n=2$ we find
\ba
\d_{j_2k_2}\d_{j_1k_1} \d^g_{\Ld}\x_N
&=& \Big( \d^g_{\Ld_{j_2k_2}(\Ld_{j_1k_1}(\Ld))} 
+\d^g_{\Ld_{j_1k_1}(\Ld)}\d_{j_2k_2}     
\nn\\&&\quad
+ \d^g_\Ld \d_{j_2k_2}   \d_{j_1k_1}  
+\d^g_{\Ld_{j_2k_2}(\Ld)} \d_{j_1k_1}  \Big) \x_N\;,\qquad \x_N\in S^2_N \;.\label{hanna}
\ea
To construct the SW-map we introduce
a natural map from the fuzzy sphere into the commutative sphere\footnote{clearly, the fuzzy sphere is here mapped into the function algebra of the commutive sphere, which we shall, however, also denote $S^2$.} 
\ba
\t: S^2_N\rightarrow S^2\;.
\ea
The map is constructed by simply replacing the generators $\xh_i$ in the expansion 
\ba
\x_N =\sum_{a_1a_2a_3}c_{a_1a_2a_3}\xh_1^{a_1}\xh_2^{a_2}\xh_3^{a_3}\;,\qquad \x_N\in S^2_N\;,c_{a_1a_2a_3}\in \mathbb{C}\;,
\ea
with the coordinates $x_i$ of $S^2$
\ba
\t(\x_N) = \sum_{a_1a_2a_3}c_{a_1a_2a_3}x_1^{a_1}x_2^{a_2}x_3^{a_3}\;.\label{projection}
\ea
This is a uniquely defined map. We denote 
\ba
\t(\F_N):=\f\;,
\qquad
\t(A_i):=a_i \;,
\ea
where $\f$ and $a_i$ are the corresponding fields on $S^2$. Further, we denote by
\ba
\t(\Ld):=\l_0\;,
\ea
the commutative gauge parameter corresponding to $\Ld$.
Next, we define the map between the fuzzy sphere and the function algebra of the commutative sphere
\ba
\r: S^2_N \rightarrow S^2\;,
\ea
as a formal Taylor expansion
\ba
\r(\x_N)= \sum_{n=0}^\infty \frac{1}{n!}\a^n \e_{i_1 j_1 k_1}x_{k_1}\ldots\e_{i_n j_n k_n}x_{k_n}  \t(\d_{i_1 j_1}\ldots \d_{i_n j_n}\x_N)\;,\quad\x_N\in S^2_N \;.  \label{SWSW}
\ea
In case $\d_{ij}\x_N=0$ we simply have
\ba
\r(\x_N) = \t(\x_N)\;.
\ea
Since neither $\cx_{ij}$ in (\ref{lalalaja}) nor $\cx_{ijk}$ in (\ref{ttt}) diverge in the commutative limit
this formal Taylor expansion is well defined for both the scalar field $\F_N$ and the gauge field $A_i$. If we insert $\F_N$ in (\ref{hanna}), generalize it to any order $n$ and set $\a=0$ in the sense of (\ref{projection}), we finally find the relation for the complex scalar field
\ba
\r\big( \d^g_\Ld \F_N \big)  = \d^g_{\l}\r\big(\F_N  \big) \label{jau}\;.
\ea
Repeating the steps for the gauge field we find the relation 
\ba
\r\big( \d^g_\Ld A_i \big)  = \d^g_{\l}\r\big(A_i  \big) \label{jau1}\;.
\ea
In (\ref{jau}) and (\ref{jau1}) the commutative gauge parameter $\l[\Ld,a,\q]$ is related to the noncommutative gauge parameter $\Ld$ through the map
\ba
\l[\Ld,a,\a]&=& \l_0 + \a\e_{j_1k_1l_1}x_{l_1}\t(\Ld_{j_1k_1}(\Ld)) 
\nonumber\\&&
+ \frac{\a^2}{2}\e_{j_2k_2l_2}x_{l_2}\e_{j_1k_1l_1}x_{l_1}\t(\Ld_{j_1k_1}(\Ld_{j_2k_2}(\Ld))) + \ldots\;.
\label{ethiopien}
\ea 
The relations (\ref{jau}) and (\ref{jau1}) are the gauge equivalence relations on the fuzzy sphere for the scalar field and the gauge field respectively, and $\r(\F_N)$ and $\r(A_i)$ are the SW-maps of the scalar field and the gauge field. 

To first order the SW-map $\r(\F_N)$ for the scalar field reads
\ba
\r(\F_N) = \f + \a\e_{ijk}x_k \t(\cx_{ij}) + \co(\a^2)\;,\label{THEMAP}
\ea
where $\cx_{ij}$ is given in (\ref{lalalaja}). 

The above result for the scalar field is of course slightly peculiar because we chose to analyze a field in the adjoint representation: The field's gauge transformation vanishes in the commutative limit and thus renders the commutative 'gauge' theory obsolete. Nevertheless, the analysis may easily be repeated for (scalar) fields in the (anti-) fundamental representation and secondly, one can in a straight forward manner extend the analysis to the non-Abelian case. In fact, our analysis does not depend on the 'internal' structure of the fields in question, only on their spin. 
In this light let us check -- up to first order -- that this map does indeed give a gauge-covariant map. We first calculate
\ba
\r(\d_\Ld^g\F_N) = \r({\rm{i}}[\Ld,\F_N]) = \frac{2}{R^2} \a\e_{ijk}x_k(L_i\l_0)( L_j\f) + \co(\a^2)\;,
\ea
and then
\ba
\d_\l^g\r(\F_N) &=& \d_\l^g(\f + \a\e_{ijk}x_k \t(\cx_{ij}))+ \co(\a^2) 
\nn\\
&=& \frac{2}{R^2} \a\e_{ijk} x_k (L_i\l_0)( L_j\f) + \co(\a^2)\;,
\ea
confirming (\ref{jau}) to first order in $\a$.

For the gauge field the SW-map $\r(A_i)$ reads to first order
\ba
\r(A_i) &=& a_i + \a\e_{jkl}x_j \t(\cx_{ikl}) + \co(\a^2)
\nn\\&=& a_i + \frac{\a}{ R}\e_{jkl}x_l a_k\big(
{\rm{i}}  J^{\mbox{\tiny cl}}_l a_i + f_{li} 
\big)   + \co(\a^2) \label{sadd}\;,
\ea
where $\cx_{ijk}$ is given in (\ref{ttt}) and 
\ba
f_{li}={\rm{i}}  J^{\mbox{\tiny cl}}_l a_i - {\rm{i}}  J^{\mbox{\tiny cl}}_i a_j +\e_{lim}a_m \;,
\ea
denotes the Abelian field strength on $S^2$. In (\ref{sadd}) we used the commutative constraint $a_i x_i=0$ to remove otherwise divergent terms. The expansion can again be checked to fulfill (\ref{jau}) order by order. The first order term in (\ref{sadd}) agrees algebraically with the result found in \cite{Hayasaka:2002db}. Clearly, higher order terms can be computed in a straight forward manner.\\

\subsection{The gauge parameters}

Until now we have not specified how the gauge parameter $\Ld$ should transform under $\d_{ij}$. First, if we choose
\ba
\d_{ij}\Ld = \frac{1}{4\a R}\Big(\{A_i,[\Ld,\xh_j]\}- \{A_j,[\Ld,\xh_i]\}\Big) \;,
\ea
we obviously find that $\Ld_{ij}(\Ld)=0$ in (\ref{gpa}). This choice leads to $\l[\Ld,a,\a]=\l_0$ in (\ref{ethiopien}) and yields a SW-map of the gauge parameter
\ba
\r(\Ld) = \l_0 + \a\e_{ijk}x_k\t(\d_{ij}\Ld)   + \co(\a^2)\;,
\ea
which in the Moyal limit gives the condition on the noncommutative gauge parameter usually found in the literature. In the general case where we do not specify any $\d_{ij}$ transformation of $\Ld$ we may define $a_\a =\tfrac{\a}{2R}\e_{ijk} x_k a_i L_j$ as well as the derivation $\d_\a=\a\e_{ijk}x_k\d_{ij}$ and write
\ba
\l[\Ld,a,\a]= \sum_{n=0}^\infty\frac{1}{(n+1)!}\t((a_\a + \d_\a)^n\Ld)\;. \label{res1}
\ea
In this notation the SW-map $\r(\F_N)$ takes the form
\ba
\r(\F_N)[\f,a,\a]= \sum_{n=0}^\infty\frac{1}{(n+1)!}\t(\d_\a^n \F_N)\;,\label{res2}
\ea
as well as the SW-map for the gauge field
\ba
\r(A_i)[a,\a]= \sum_{n=0}^\infty\frac{1}{(n+1)!}\t(\d_\a^n A_i)\;.\label{res3}
\ea
Equations, (\ref{res1}), (\ref{res2}), and (\ref{res3}) resemble very much the general results found in \cite{Jurco:2000fb,Jurco:2000fs,Jurco:2001my,Jurco:2001kp}.

\subsection{Different choices of deformation parameters}

Apart from the option of adding covaritant terms to $\cx$ and $\cx_i$ in (\ref{lalalala}) and (\ref{rel3}) (and thereby changing $\d_{ij}$) there appears to be another freedom in constructing a SW-map -- as it is defined in section \ref{11} -- on the fuzzy sphere.
First, notice that the gauge equivalence condition (\ref{jau}) is also fulfilled if we simply choose the trivial expansion
\ba
\d_{ij}\F_N=0\;,\quad \d_{ij}(\x_N\z_N)=0 \;, \qquad \x_N,\z_N\in S_N^2\;. \label{trivial}
\ea
In fact, if we write
\ba
\r^\prime(\F_N)= \t(\F_N) + \g_{ij}\t(\d_{ij}\F_N ) + \co(\g_{ij}^2)\;,
\ea
for some deformation parameter $\g_{ij}$, then the SW-map described by (\ref{SWSW}) amounts to the choice $\g_{ij}=\a\e_{ijk}x_k$. This is a sound choice because $\g_{ij}$ is not gauge-transformed. Another possible choice is of course $\g_{ij}=0$ which amounts to the trivial SW-map described by (\ref{trivial}). In principle, any choice of a (`small') gauge-invariant, antisymmetric $\g_{ij}$ of dimension [length]$^2$ will lead to a map fulfilling the gauge equivalence condition (\ref{jau}). 
Of course, there are not very many ways to construct such a $\g_{ij}$ using elements already present in the model. However, if we introduce a constant unit vector $\bar{n}_i$ we could write $\g_{ij}$ as its dual: 
\ba
\g_{ij}= \a R\e_{ijk}\bar{n}_k\;.
\ea   
This leads to a whole new class of non-trivial maps\footnote{These maps, however, do not represent deformations in the sense of deformation quantization where the deformation parameter must be $\a\e_{ijk}x_k$.} fulfilling the gauge equivalence condition (\ref{jau}). 

These considerations demonstrate first of all that there is a freedom in the choice of $\g_{ij}$. Secondly, it becomes clear that the first requirement given in section \ref{11} regarding the SW-map as an expansion is a necessary condition. For the trivial choice (\ref{trivial}) there is no deformation of the commutative theory.

\subsection{A consistent deformation of gauge theories on the classical sphere}\label{100}

The SW-map applied to the full action leads to a deformation of classical gauge theory on the sphere in a manner consistent with a gauge theory on the fuzzy sphere. The $n$'th order deformation
\ba
\r(S[A,\Psi]) =  S[a,\psi] + \sum_{i=1}^n\a^i S_i[a,\psi]  + \co(\a^{n+1})\,,\label{thecomact}
\ea
where $S[a,\psi]$ is the action of ordinary gauge theory on the sphere and $S_i[a,\psi]$ are higher order terms, gives a classical action on the ordinary sphere which contains information about the noncommutative structure of the fuzzy sphere. Further -- as a check of consistency -- it is easily found that the SW-map preserves the gauge constraint (\ref{hovsa})
\ba
\r\big(\xh_i A_i + A_i\xh_i +\a R A_i A_i \big) = 2a_i x_i  \;,
\ea
as well as the spherical constraint
\ba
\r\big(\xh_i \xh_i \big) = x_i x_i\;.
\ea
It is, however, important to realize that the action (\ref{thecomact}) is an object very different from the actions on the fuzzy sphere, (\ref{ac2}) and (\ref{ac3}). The inherent non-locality of the fuzzy sphere is lost together with the truncation of the function algebra. The expansion (\ref{thecomact}) does involve information about the noncommutativity -- encoded in the transformation of the product $\d_{ij}$ -- as well as information about the gauge symmetry which is encoded in the SW-map of the fields. But these maps -- seen from the commutative sphere point of view -- represent little more than field redefinitions (see also \cite{Grimstrup:2002af}). Thus, it is not clear to us to what extent the action (\ref{thecomact}) -- or any other commutative action obtained via a SW-map -- captures the physics of the noncommutative algebra from where it originates.

\section{Covariant maps between different fuzzy spheres}\label{123}

On noncommutative $\mathbb{R}^4_\q$ the SW differential equation leads to the SW-map between the noncommutative and commutative (function) algebras. One can, however, equally well map gauge theories living on {\it different} $\mathbb{R}^4_\q$'s characterized by different $\q$'s, say, $\q_1$ and $\q_2$. In this case, the deformation parameter is $\d\q=\q_2-\q_1$. Basically, this is possible because the noncommutativity parameter $\q_{ij}$ lies in the center of the algebra.

On the fuzzy sphere the noncommutativity parameters do not lie in the center of the algebras and therefore it is not possible to generalize the SW-map found in the previous section to a map between different fuzzy spheres. 
However, if we permit maps involving a constant (unit) vector it is possible to construct a map from $S^2_N$ to $S^2_{N^\prime}$ as long as $N^\prime$ is greater than $N$. 

We denote by
\ba
\a_N=\frac{R}{\sqrt{\tfrac{N}{2}(\tfrac{N}{2}+1)}}\;, \quad 
\ea
the noncommutativity parameter of the fuzzy sphere $S^2_N$ and write $J_i^{\mbox{\tiny $N$}}$ and $\xh_i^{\mbox{\tiny $N$}}$, $i\in\{1,2,3\}$, for the corresponding generators (recall: $\a J_i^{\mbox{\tiny $N$}}=\xh_i^{\mbox{\tiny $N$}}$). Again, our starting point is the relation (\ref{hanna}) generalized to any order $n$. We first define a mapping $\s$ between $S^2_N$ and $S^2_{N^\prime}$ by identifying the generators $J_i^{\mbox{\tiny $N$}}$ on $S^2_N$ with the generators $J_i^{\mbox{\tiny $N$}}$ on $S^2_{N^\prime}$
\ba
\s(J_i^{\mbox{\tiny $N$}}):=J_i^{\mbox{\tiny $N^\prime$}}\;.
\ea
In this way a general element on the fuzzy sphere $S^2_N$
\ba
\x_N  = \sum_{a_1a_2a_3}c_{a_1a_2a_3}(J^{\mbox{\tiny $N$}}_1)^{a_1}(J^{\mbox{\tiny $N$}}_2)^{a_2}(J^{\mbox{\tiny $N$}}_3)^{a_3}\;,
\ea
is mapped into an element on the fuzzy sphere $S^2_{N^\prime}$:
\ba
\s(\x_N)  := \sum_{a_1a_2a_3}c_{a_1a_2a_3}(J^{\mbox{\tiny $N^\prime$}}_1)^{a_1}(J^{\mbox{\tiny $N^\prime$}}_2)^{a_2}(J^{\mbox{\tiny $N^\prime$}}_3)^{a_3}\;.
\ea
and define the map
\ba
\tilde{\r}: S_N^2\rightarrow S_{N^\prime}^2
\ea
by
\ba
\tilde{\r}(\x_N)=\sum_{n=0}^\infty \frac{1}{n!}\g_{i_1 j_1}\ldots\g_{i_n j_n } \s(\d_{i_1 j_1}\ldots \d_{i_n j_n}\x_N )\;,
\ea
where $\x_N\in S^2_N$ and $\g_{ij}\in S^2_{N^\prime}$ . For $\tilde{\r}$ to fulfill a gauge equivalence condition like (\ref{jau}) we need $\g_{ij}$ not only to be gauge invariant, antisymmetric and of dimension [length]$^2$ but also to lie in the center of the algebra $S^2_{N^\prime}$. This means, in particular, that we {\it cannot} choose $\g_{ij}$ to
\ba
\g_{ij}\not\sim\e_{ijk}\xh_k^{\mbox{\tiny $N^\prime$}}\;,
\ea
since this obviously does not satisfy the above criteria. However, as discussed in the previous section, if we allow ourselves to introduce a constant unit vector $\bar{n}$ we are lead to a large class of maps
\ba
\g_{ij}=(\a_{N^\prime}-\a_N)R\e_{ijk}\bar{n}_k\;.
\ea
For the scalar field, this map reads to first order
\ba
\tilde{\r}(\F_N)=\s(\F_N) + (\a_{N^\prime}-\a_N)R\e_{ijk}\bar{n}_k \s(\cx_{ij}) + \co((\a_{N^\prime}-\a_N)^2)\;.
\ea
The presence of an antisymmetric tensor $\g_{ij}$ breaks rotational invariance on the fuzzy sphere. Two different maps between the fuzzy spheres $S^2_N$ and $S^2_{N^\prime}$ characterized by unit vectors $\bar{n}$ and $\bar{n}^\prime$ will, however, be related through the rotation mapping $\bar{n}$ into $\bar{n}^\prime$.

\section{Summary and conclusion}\label{9}

In the present paper we have constructed covariant coordinate transformations on the fuzzy sphere. These transformations, which yield ordinary covariant coordinate transformations in the commutative limit, were used to construct a Seiberg-Witten map on the fuzzy sphere for $U(1)$ gauge theories. Seiberg-Witten maps for both a complex scalar field and a gauge field were constructed as formal power-series which permit computation of the map to arbitrary order. The map was demonstrated to coincide with the Seiberg-Witten map on the Moyal plane in the appropriate limit. 
Furthermore, we find that the setup cannot be generalized in a straight forward manner to obtain a covariant mapping between different fuzzy spheres (as is the case on noncommutative $\mathbb{R}^4_\q$). By introducing a constant vector can one construct such mappings.

The existence of the Seiberg-Witten map on the fuzzy sphere was expected on grounds of the general work by Jurco et al. \cite{Jurco:2001my}. From \cite{Jurco:2001my} it was, however, less clear how the map was to be constructed since the approach involves a smooth deformation of the noncommutativity parameter, which is not applicable on the fuzzy sphere where the noncommutativity parameter $\a$ is discrete. Furthermore, we have demonstrated that the approach using covariant coordinate transformations \cite{Bichl:2001yf} -- which applies to any field in any representation of the gauge group -- can be applied to fuzzy spheres.

The analysis presented in this paper takes place in the algebra and is independent of the star-product. We believe that this gives a clearer view of what a Seiberg-Witten map represents and what it might be good for. It turns out that the Seiberg-Witten map is a map from the noncommutative algebra into the commutative algebra rather than an expansion of the noncommutative field theory in terms of the noncommutativity parameter and commutative fields. 


We believe that the approach leading to the SW-map on the fuzzy sphere, which was presented in this paper, can be applied also to other noncommutative algebras, for example the noncommutive torus. It would be interesting to investigate whether the SW-map can be constructed non-perturbatively for simple algebras such as the fuzzy sphere and the noncommutative torus.

\acknowledgments

We are grateful to Branislav Jurco and Peter Schupp for helpful clarifications on their work and to Raimar Wulkenhaar for reading the manuscript and for providing numerous valuable comments. This research was partly supported by TMR grant no. HPRN-CT-1999-00161 and by grants from the Icelandic Research Council and
the University of Iceland Research Fund.

\section*{Appendices}
\addcontentsline{toc}{section}{Appendices}

\subsection*{Appendix A: The derivation of (\ref{INIT})} \label{a1}


\def\theequation{A.\arabic{equation}}
\setcounter{equation}{0}
\label{apple}

We first solve (\ref{INIT}). For notational simplicity we write $\tA_i=\a R A_i$.
To bring the dilational term in (\ref{INIT}) into covariant form we write
\ba
\frac{{\rm{i}}\e_i}{2R\a}\e_{ijk}\{\xh_j,[\xh_k,\F_N]\}&=& \frac{{\rm{i}}\e_i}{2R\a}\e_{ijk}\{X_j,[X_k,\F_N]\} -\frac{{\rm{i}}\e_i}{2R\a}\e_{ijk}\{\ti{A}_j,[\xh_k,\F_N]\}
\nn\\&&
-\frac{{\rm{i}}\e_i}{2R\a}\e_{ijk}\{x_j,[\ti{A}_k,\F_N]\}-\frac{{\rm{i}}\e_i}{2R\a}\e_{ijk}\{\ti{A}_j,[\ti{A}_k,\F_N]\}\;,
\ea
after which (\ref{INIT}) obtains the form
\ba
\OO + {\rm{i}}[\tilde{\Ld},\F_N] &=& \frac{{\rm{i}}\e_i}{2R\a}\e_{ijk}\{X_j,[X_k,\F_N]\} -\frac{{\rm{i}}\e_i}{R\a}\e_{ijk}\{\ti{A}_j,[\xh_k,\F_N]\}
\nn\\&&
-\frac{{\rm{i}}\e_i}{2R\a}\e_{ijk}[\{\xh_j,\ti{A}_k\},\F_N] - \frac{{\rm{i}}\e_i}{2R\a}\e_{ijk}\{\ti{A}_j,[\ti{A}_k,\F_N]\} 
+ \cx\;.
\ea
Now, we set
\ba
\tilde{\Ld} =\frac{\e_i}{2R\a}\e_{ijk}\{\xh_j,\ti{A}_k\} \;, \label{sgu}
\ea
 and 
\ba
\OO^\prime =\OO -\frac{{\rm{i}}\e_i}{2R\a}\e_{ijk}\{X_j,[X_k,\F_N]\}\;,
\ea
where $\OO^\prime(\F_N)$ is an arbitrary covariant term that must vanish in the commutative limit. In terms of $A$ we find
\ba
\cx &=& R\a\e_i\e_{ijk}\Big(
 \frac{{\rm{i}}}{R\a}\{A_j,[\xh_k,\F_N]\}
+ \frac{{\rm{i}}}{2}\{A_j,[A_k,\F_N]\}  \Big)  \label{thee}\;.
\ea

\subsection*{Appendix B: The solution to (\ref{saddd})}
\def\theequation{B.\arabic{equation}}
\setcounter{equation}{0}
\label{apple1}

Again, we proceed by noting that the full solution consists of a gauge transformation and a covariant part
\ba
\tilde{\d}_\e A_i &=&
{\rm{i}}\frac{\e_j}{2R\a}\e_{jkl}\{\xh_k,[\xh_l,A_i]\} 
\nn\\
&&+ \frac{ \e_j}{2R}\Big(\{ \xh_k,A_k\}\d_{ij} + \{\xh_j,A_i\} - 2\{\xh_i,A_j\}  \Big) + \cx_i 
\nn \\
&=& {\rm{i}}\frac{1}{R\a}[X_i,\tilde{\Ld}] + \OO_i\;. \label{comparing}
\ea
Let us start by writing down the covariant version of $\d_\e A_i$ 
\ba
&&{\rm{i}}\frac{\e_j}{2(R\a)^2}\e_{jkl}\{X_k,[X_l,X_i]\} 
+ \frac{ \e_j}{2R^2\a}\Big(- \{X_j, X_i \}  +\d_{ij}\{X_k, X_k\}\Big)
\nn\\ &=&  
{\rm{i}}\frac{\e_j}{2(R\a)^2}\e_{jkl}\{X_k,[X_l,X_i]\} 
+ \frac{ \e_j}{2R^2\a}\Big(-\{X_j, X_i \}  +2\d_{ji}R^2\Big)
\;.
\ea
We find (again: $\tilde{A}_i=\a R A_i$)
\ba
{\rm{i}}\frac{\e_j}{2(R\a)^2}\e_{jkl}\{\xh_k,[\xh_l,\tA_i]\} 
+ \frac{ \e_j}{2R^2\a}\Big(\{ \xh_k,\tA_k\}\d_{ij} + \{\xh_j,\tA_i\}- 2\{\xh_i,\tA_j\}\Big)
\hspace{-10cm}\nn\\
&=&
{\rm{i}} \frac{\e_j}{2(R\a)^2}\e_{jkl}\{X_k,[X_l,X_i]\} + \frac{ \e_j}{2R^2\a}\Big(-\{X_j, X_i \}  +2\d_{ji}R^2\Big)
\nn\\&&
-{\rm{i}} \frac{\e_j}{2(R\a)^2}\e_{jkl}[\{ \tA_k,\xh_l\}  ,X_i]  
-2{\rm{i}} \frac{\e_j}{2(R\a)^2}\e_{jkl}\{ \tA_k,[\xh_l,\tA_i]\}   
\nn\\&&
-{\rm{i}} \frac{ \e_j}{2(R\a)^2}\e_{jkl}\{ \tA_k,[ \tA_l,X_i]\} 
\nn\\&&
-\frac{ \e_j}{2R^2\a}\e_{jkl}\Big( 
\e_{lim}\{\tA_k,\tA_m\}
-\e_{lim}\{\xh_k,\tA_m\}
\Big)\;.
\ea
Comparing this with (\ref{comparing}) we read off (we have already set the possible arbitrary covariant term to zero)
\ba
\OO_i = {\rm{i}} \frac{\e_j}{2(R\a)^2}\e_{jkl}\{X_k,[X_l,X_i]\} + \frac{ \e_j}{2R}\Big(-\{X_j, X_i \}  +2\d_{ji}R^2\Big)\;,
\ea
as well as the gauge parameter
\ba
\tilde{\Ld}=  \frac{\e_j}{2}\e_{jkl}\{\xh_k,  A_l\}\;,
\ea
which is the same as found in (\ref{sgu}). Finally we find
\ba
\cx_i &=& 
\a R \e_j\e_{jkl}\Big(
{\rm{i}} \frac{1}{\a R}\{ A_k,[\xh_l,A_i]\}   
+{\rm{i}} \frac{ 1}{2\a R}\{ A_k,[ A_l,X_i]\} 
\nn\\&&
+\frac{ 1}{2 R}
\e_{lim}\{A_k,A_m\}
-\frac{1}{2\a R^2}\e_{lim}\{\xh_k,A_m\}\Big)\;.
\ea

\bibliographystyle{JHEP}

\end{document}